# Identification of Helicopter Dynamics based on Flight Data using Nature Inspired Techniques

S.N. Omkar[1], Dheevatsa Mudigere[2], Senthilnath J[1], M. Vijaya Kumar[3]

[1]Department of Aerospace Engineering, Indian Institute of Science-Bangalore, 560 012, KA, India

[2]Lehrsthul für Scientific Computing in Computer Science, Informatik Department, TUM, München – 85478.

[3] Rotary Wing Research and Design Center, HAL, Bangalore – 560017.

Email:omkar@aero.iisc.ernet.in

*Abstract*— The complexity of helicopter flight dynamics makes modeling and helicopter system identification a very difficult task. Most of the traditional techniques require a model structure to be defined a priori and in case of helicopter dynamics, this is difficult due to its complexity and the interplay between various subsystems. To overcome this difficulty, non-parametric approaches are commonly adopted for helicopter system identification. Artificial Neural Network are a widely used class of algorithms for non-parametric system identification, among them, the Nonlinear Auto Regressive eXogeneous input network (NARX) model is very popular, but it also necessitates some in-depth knowledge regarding the system being modelled. There have been many approaches proposed to circumvent this and yet still retain the advantageous characteristics. In this paper we carry out an extensive study of one such newly proposed approach - using a modified NARX model with a two-tiered, externally driven recurrent neural network architecture. This is coupled with an outer optimization routine for evolving the order of the system. This generic architecture is comprehensively explored to ascertain its usability and critically asses it's potential. Different instantiations of this architecture, based on nature inspired computational techniques (Artificial Bee Colony, Artificial Immune System and Particle Swarm Optimization) are evaluated and critically compared in this paper. Simulations have been carried out for identifying the longitudinally uncoupled dynamics. Results of identification indicate a quite close correlation between the actual and the predicted response of the helicopter for all the models.

## I. INTRODUCTION

Helicopter system identification is the extraction of system characteristics/dynamics from measured flight test data (Maine, R. E., & Iliff, K. W., 1985; Anon, 1991). The complexity of helicopter flight dynamics makes, modelling and helicopter system identification a very challenging task. Unlike fixed-wing aircrafts, the helicopters exhibits a high degree of inter-axis coupling, highly unstable, non-minimum phase dynamic characteristics and large response variations with flight condition. These characteristics of the helicopter make it a highly non-linear and a complex dynamical system. Further, the prediction of aeromechanical forces, loads on the rotor system and main rotor wake interferences with the empennage and tail rotor require wind tunnel experiments and flight tests. But the wind tunnel experimental data suffers from scale effects and model deficiencies. Therefore, a key tool for helicopter flight/ground test correlation is provided by system identification using flight data.

Identification of a system requires picking a function (or model) so as to approximate the input-output behaviour of the system (the helicopter in this case) in the "best" possible manner. There has been considerable amount of work carried out in this regard, exploring the various methods available for identification of dynamical systems (Miller, W. T., Sutton, R. S., & Werbos, P.J. 1990;

Narendra K. S. & Parthasarathy, K., 1989; Narendra K. S. & Parthasarathy, K., 1990; Narendra K. S. & Parthasarathy, K., 1991; Ichikawa. Y & Sawa. T, 1992; Sastry, P. S., Santharam G. & Unnikrishnan K. P., 1994; Chen, S., Billings, S. A., & Grant, P. M., 1990; Hoskins, D. A., Hwang, J. N., & Vagners J., 1992). Identification of nonlinear physical models continues to be a challenge since both the structure and parameters of the physical model must be determined. Many existing system identification methods are based on parametric identification. Structure determination is based on expert knowledge of the underlying physics, lacking which often trial and error approach to test candidate model structures is employed. Possible structures are deduced from engineering knowledge of the system and the parameters of these models are estimated. But in the case of a helicopter, defining an a priori model is difficult due to interaction between the various subsystems like the rotor, fuselage, power plant, tail rotor and transmission systems (Tischler, M. B., 1996) the dynamics are of relatively higher order, and it is difficult to know how many states to include and which states are important. Also, increase in the nonlinearity, uncertainty and complexities of the model together with the stringent specifications of accuracy limits to be maintained renders modelling helicopter systems a daunting task. This initiated an interest among researchers to identify the system characteristics using nonparametric methods.

Artificial Neural Network (ANN) have found widespread application in nonlinear dynamic system identification as universal approximators (Miller, W. T., Sutton, R. S., & Werbos, P.J. 1990; Narendra K. S. & Parthasarathy, K., 1989; Narendra K. S. & Parthasarathy, K., 1990; Narendra K. S. & Parthasarathy, K., 1991; Ichikawa. Y & Sawa. T, 1992; Sastry, P. S., Santharam G. & Unnikrishnan K. P., 1994; Chen, S., Billings, S. A., & Grant, P. M., 1990; Hoskins, D. A., Hwang, J. N., & Vagners J., 1992). ANNs have also been used for helicopter system identification, Vijaya Kumar et.al. (2003) have explored the different Recurrent Neural Networks (RNN) for the identification of helicopter dynamics and based on the results, the practical utility, advantages and limitations, the models have been critically appraised. The authors after a comprehensive study of the three popular RNN architecture – Nonlinear Auto Regressive eXogenous (NARX) input model, Memory Neuron Network (MNN) model and Recurrent Multi-Layer Perceptron (RMLP) model, concluded that the NARX model is most suitable for the identification of helicopter dynamics (Vijay Kumar, M. et. al. 2003). The NARX model, proposed by Narendra (1989, 1990 & 1991) uses Tapped-Delay-Lines (TDL) and the Multi-Layer Perceptron neural network architecture (MLP) for non-linear system identification. The NARX model is trained using the Back Propagation (BP) algorithm. Although NARX model is popular and widely used, it has certain drawbacks. NARX model requires the necessary past inputs and outputs of the system being modelled, to be fed as explicit inputs to the network; this necessitates a fairly in-depth knowledge of the system. But, as previously discussed, in case of helicopters this is challenging ask. While, in principle, we can always feed a "sufficient" number of past values to the network, in practice, all reported applications assume that the exact order is known. Both from aesthetics and practicality, learning transformations in this manner may not lead to versatile dynamical models (Sastry, P. S., Santharam G. & Unnikrishnan K. P., 1994). This can be overcome by using alternatives such as the memory neuron networks (MNN) (Sastry, P. S., Santharam G. & Unnikrishnan K. P., 1994) etc... But The NARX model has been proven to have far superior stability characteristics when compared to MNN and the other models (Vijay Kumar, M. et. al. 2003). This makes the NARX model a very popular choice for identification and adaptive control of dynamical systems even with its set of limitations. Mudigere, D., et.al (2008) have proposed a PSO (particular the particle swarm optimization) (Eberchart, R., & Kennedy, J., 1995) Driven Recurrent Neural Network (PSO-NARX) Model, for circumventing this draw-back of the conventional NARX model. This model employs a PSO based algorithm for evolving the order (and relative degree) of the system, coupled with the (NARX-like) multi-layer perceptron recurrent neural network with tapped delay lines, trained by a PSO based learning algorithm, for the identification of the dynamical system. This model employs a generic II-Tier architecture, which offers a great deal of flexibility and allows for using a host of different algorithm combinations for evolving the order and training the system.

The objective of the current work is to further develop this concept by comprehensively exploring this generic II-Tier architecture with different combinations of nature inspired techniques for evolving the order and training the network. In order to exploit the inherit flexibility of this architecture to develop an efficient and accurate model for identification of non-linear dynamical systems. In this paper we consider algorithms inspired from nature to achieve the above objective – in particular the particle swarm optimization (PSO) (Eberchart, R., & Kennedy, J., 1995), Artificial Bee Colony (ABC) (Karaboga, D., & Basturk, B., 2008; Karaboga, D., Akay, B., & Ozturk, C., 2007) and Artificial Immune System (AIS) (De Castro, L. N., & Von Zuben, F. J., 2000; Dasgupta D., 1999) are considered. This results in three different variants of this model for nonlinear dynamical system identification, which do not necessitate any apriori information about the system and the favourable characteristics (stability) of the popular NARX model. Further the proposed models are successfully employed for the identification of helicopter dynamics using flight test data. These different models are critically appraised and compared.

## II. NATURE INSPIRED TECHNIQUES

Nature inspired technique is the field of research that works with computational techniques inspired in part by nature and natural systems. These nature inspired techniques provide a more robust and efficient approach for solving complex real-world problems (Bäck, T., & Schwefel, H. P., 1993). Many nature inspired techniques such as Artificial Bee Colony (ABC) (Karaboga, D., & Basturk, B., 2008), Artificial Immune System (AIS) (De Castro, L. N., & Von Zuben, F. J., 2000), Artificial Neural Network (Haykin, S., 1994), Genetic Algorithm (GA) (Goldberg, D. E., 1989), Particle Swarm Optimization (PSO) (Eberchart, R., & Kennedy, J., 1995) etc.. have been proposed. Since they are heuristic and stochastic in nature, they are less likely to get stuck in local minimum, and they are based on populations made up of individuals with a specified behavior similar to biological phenomenon. These characteristics led to the development of nature inspired computation as it is increasingly applied in various domains (Engineering problems). Presently, it is one of the important areas of research.

### A. Artificial Bee Colony

Artificial Bee Colony (ABC) (Karaboga, D., & Basturk, B., 2008) is a class of optimizing numerical problem based on swarm intelligence, investigating the foraging behaviour of bees. In

ABC algorithm, the colony of artificial bees contains three groups of bees which include scout bees, employed bees and onlookers. A bee carrying out random search is called a scout. A bee waiting on the dance area for making decision to choose a food source is called an onlooker and a bee going to the food source visited by itself previously is named an employed bee.

At the first step, create a population of 'n' artificial bees placed randomly in the search space representing the food source position. After initialization, the population of the positions (solutions) is subjected to repeated iteration of the search processes of the employed bees, the onlooker bees and scout bees.

For each solution $x_{ij}$, where $i = 1, 2...n$ and $j$ is dimensional vector. The scout bees explore a new food source with $x_i$. This operation can be defined as in (1)

$$x_i^j = x_{min}^j + (x_{max}^j - x_{min}^j) * rand(0,1) \tag{1}$$

The population spread is restricted within the search space $S$ i.e $x_{ij} \in S$ and in the equation (1) $x_{min}$ and $x_{max}$ are the lower and upper limit respectively of the search scope along each dimension;

The bees which have explored the food source are selected as employed bees. Which results in a modification on the position (solution) in those candidate bees' memory, as a function of the local information and tests the nectar amount (fitness value) of the new source. After all the employed bees complete the search process; they communicate the nectar information of the food sources and their position information with the onlooker bees in the dance area. An onlooker bee evaluates the nectar information taken from all employed bees and chooses a food source with better nectar amount.

An artificial onlooker bee chooses a food source depending on the new positions, using the equation (2).

$$P_i = \begin{cases} v_i, & if\ (f(x_i) \geq f(v_i)) \\ x_i, & if\ (f(x_i) \leq f(v_i)) \end{cases} \tag{2}$$

In order to select the better nectar position found by an onlooker, $O_b$ is defined as

$$O_b = \arg\min_{P_i} f(P_i),\ 1 \leq i \leq n \tag{3}$$

Where $P_i$ is the best fitness value of the solution $i$ which is proportional to the nectar amount of the food source in the position $i$ and $n$ is the number of food sources.

In order to produce a candidate food position from the old one in memory, the ABC uses the following equation (4):

$$v_{ij} = x_{ij} + \gamma(x_{ij} - x_{kj}) \tag{4}$$

where $k=1, 2,..., n$ and $j = 1, 2,...,D$ are randomly chosen indices. Although $k$ is determined randomly, it has to be different from $i$. $\gamma$ is an adaptively generated random number which controls the learning/adoption rate.

### B. Artificial Immune System

As described by De Castro, et.al (2002), Artificial Immune Systems (AIS) are adaptive systems inspired by theoretical immunology and observed immune functions, principles and models which are applied to problem solving.

Much of the early work carried out in the development of AIS has been similar to genetic and evolutionary computation techniques. The main distinction between the field of genetic algorithms and AIS is the nature of population development. In a genetic algorithm the population is evolved using crossover and mutation (McCall, J., 2005). However in the AIS, as in evolutionary strategies reproduction is asexual (cloning), each child produced by a cell is an exact copy of its parent. Both systems then use mutation to alter the progeny of the cells and introduce genetic variation. We make use of a variant of AIS called the Clonal selection algorithm for optimization.

Clonal Selection Principle (De Castro, L. N., & Von Zuben, F. J., 2000) is one of the inspiring methodologies employed in AIS for multi objective optimization problems (Coello Coello, C.A., & Cruz Cortes, N., 2002). Based on the clonal selection principle - an algorithm is developed in which various immune system aspects are taken into account such as: maintenance of the memory cells, selection and cloning of the most stimulated cells, death of non-stimulated cells and re-selection of the clones with higher affinity and generation and maintenance of diversity.

The clonal selection algorithm works with an initial repertoire of antibodies. When an antigen is presented to it, the antibodies that are more effective in neutralizing the threat are allowed to proliferate. The least effective ones are eliminated. Among the ones selected to multiply, some amount of mutation is introduced to in the hope of finding antibodies that might perform better. However the mutation rate is lesser in antibodies with better fitness although it has a higher number of clones.

### C. Particle Swarm Optimization

Particle Swarm Optimization (PSO) is a swarm intelligence algorithm proposed by Kennedy and Eberhart in the mid 1990s. In the proposed algorithm, an agent of the swarm, called a particle, learns from the best position that it has occupied and also the best position that any particle of the swarm might have encountered. These positions are retained in memory of every particle and constantly updated to direct the swarm to the global best position. The best position of a particle is called cognitive index - *pBest*, and the best of the swarm the social index - *gBest*. The equations that govern the change in positions of a particle are:

$$V_{(i+1)} = w * V_i + \{C_p * r_1 * (pBest_i - X_i)\} + \{C_g * r_2 * (gBest - X_i)\} \quad (5)$$

$$X_{(i+1)} = X_i + V_{(i+1)} \quad (6)$$

Since each particle is constantly moving, there is a velocity vector associated with each particle that governs its motion. This is denoted by $V_i$ and $X_i$ is its position. $pBest_i$ is the best position of particle *i*. *gBest* is the global best position of the swarm. $C_p$ is the Cognitive learning rate and $C_g$ is the Social learning rate. The factors $r_1$ and $r_2$ are randomly generated within the range (0, 1) and $w$ is the inertia factor.

### III. HELICOPTER SYSTEM IDENTIFICATION – PROBLEM FORMULATION

Conventional system identification methods can be broadly classified into time-domain and frequency domain methods. In time-domain system identification, techniques such as least square estimation (Astrom, K. J., & Eykhoff, P. E., 1987), quasi-linearization (Kalaba, R., & Spingarn, K., 1982), and stochastic modelling (Kuzta, B., 1983), have been successfully used. In these methods, the model structure must be defined apriori to estimate all the required system parameters. These methods are extensively used in helicopter system identification for linear models and the approaches can be extended to nonlinear regions as well. Although time domain methods have been more frequently used, frequency domain approach is also used successfully in rotorcraft identification (Tischler, M. B., & Cauffman, M. G., 1992; Fu, K.H., & Marchand, M., 1983 & Fu, K. H., & Kaletka, J., 1990).

Modeling helicopter dynamics is a multiple input multiple output (MIMO) problem with a high degree of interaction between all the inputs and outputs.

The objective of system identification is then to construct a suitable model, such that the input-output behavior ($u(k) \rightarrow \hat{y}(k)$) of the model approximates, in some sense, the input-output behavior ($u(k) \rightarrow y(k)$) of the helicopter system, i.e. for some specified small positive constant ($\varepsilon$).

$$\|y(k) - \hat{y}(k)\| \leq \varepsilon \quad \forall\, k > 0 \qquad (7)$$

Hence, the input-output model for helicopter dynamics is described as

$$y_i(k+d) = F_i(\bar{u}(k),..\bar{u}(k-n_1+1), \bar{y}(k-1),..\bar{y}(k-n_1+1)), \quad i=1,2...,p \qquad (8)$$

where $n_1$ is the order (or equivalent delay) of the system, $d$ is the relative degree and the function $F_i(.)$ is smooth and continuous.

### IV. FLIGHT DATA ACQUISITION

Flight test is carried out in calm wind conditions on a helicopter having a soft in-plane four-bladed hinge-less main rotor and a four bladed tail rotor with conventional mechanical controls. The helicopter is instrumented to measure pitch ($q$), roll ($p$) and yaw ($r$) rates, longitudinal ($a_x$), lateral ($a_y$) and normal accelerations ($n_z$), pitch attitude ($\theta$) and roll attitude ($\phi$), indicated airspeed (V), barometric altitude (H) and sideslip ($\beta$). The four-control displacements namely, longitudinal cyclic ($\delta_{long}$), lateral cyclic ($\delta_{lat}$), collective ($\delta_{col}$) and pedal input ($\delta_{dir}$) are also measured. However, for the present study, $a_x$, $a_y$, V, H and $\beta$ are not used for simulation.

The helicopter is trimmed in straight and the pilots maintain level flight conditions at two different airspeeds of 120 Kmph and 230 Kmph. Inputs are provided manually and no automatic signal generators are used. The flight test engineer briefs pilots at the start of each flight and before carrying out each test point. For example, while giving an input manually to $\delta_{lat}$, pilot has been briefed to adjust the other control surfaces ($\delta_{long}$, $\delta_{dir}$, $\delta_{col}$) appropriately to maintain the flight condition in off-axis. Flight data for the response of the helicopter for 5 to 10% of step/doublet inputs has been recorded for a minimum of 20 sec duration.

## V. METHODOLOGY

The model proposed in this paper, is capable of identifying a nonlinear dynamical system given just a set of input-output pairs of the system under study. It does not necessitate any a priori knowledge about the system.

This model is based on the NARX model. This model employs the MLP architecture in combination with TDLs for identifying dynamical systems. As in the NARX model here also, the past values of outputs and inputs are fed back using tapped-delay-lines (TDL). The MLP approximates the nonlinear function and TDL introduces the dynamics into the model. But the distinction here is that, the order and relative degree of the system need not be known beforehand, as this model is developed to evolve by itself the order and relative degree of the system being identified. The model determines the necessary past inputs and outputs of the system being modelled that need to be fed as explicit inputs to the network. Further, the network so identified is trained using a stochastic learning algorithm based nature inspired techniques, instead of back-propagation as in the conventional NARX model. Schematic representations of the models are given in Fig. 1.

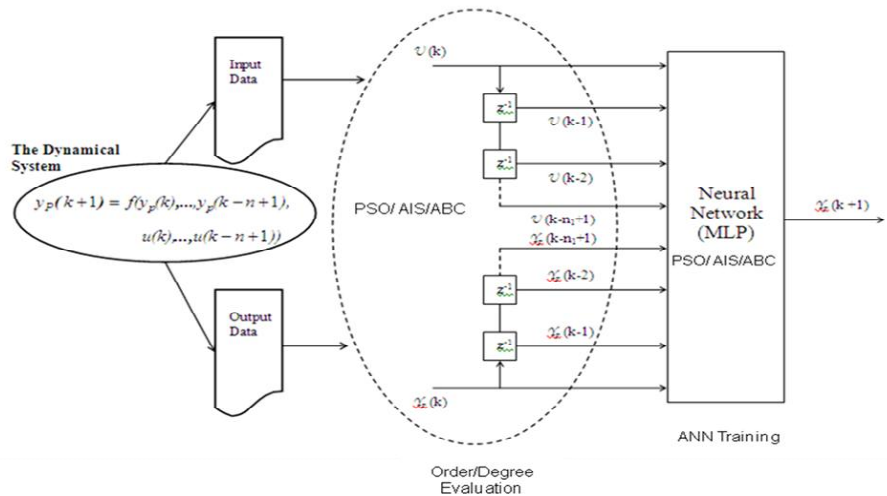

**Fig.1.** Schematic representation of the proposed modeling approach

The helicopter dynamic identification is handled in a two-tier approach. First, the order and the relative degree of the system has to be evolved i.e. the number of history values the inputs/outputs that influences the current output of the system. This determines the configuration of network as this defines the total number of parameters to be fed as explicit inputs (node in the input layer) to the network. The second step would be to train the network so identified. In this paper we employ the II-Tier architecture proposed by Mudigere, D., et.al (2008). In this paper three different nature inspired methods – ABC, AIS and PSO are used. Further 3 model variants, with different combinations of these algorithm are proposed - ABC-ABC, with I-tier ABC for determining the order of the system and a learning algorithm based on ABC for training the identified network and similarly AIS-PSO and the PSO-NARX model which used PSO based algorithms for both the tiers. These varaints use the NARX model as the basic substrate with an external ABC, PSO and AIS shell for evolving the order of the system. The generic implementation (three of the algorithm(s) used) of the above-described architecture is described below.

## A. Tier I

Tier I is for evolving the order of the system. This is to determine the number of parameters to be fed as explicit inputs to the network. Hence the decision variables in this case are the number of past values of each input and number of past values of each output – that influences the current output of the system.

Number of past values of each input is represented by, $U\ [u_1,\ u_2,\ u_3...u_i]$ and the number of past values of each output is given by, $Y\ [y_1,\ y_2,\ y_3...y_j]$, where $i$ and $j$ are the number of inputs and outputs of the system under consideration. Since the dynamics of the helicopter is represented by a MIMO system (as discussed earlier), the number of history states for each input and output has to be determined. In this module the solution-space is the various combinations of $[u_1,\ u_2,\ u_3...u_i,\ y_1,\ y_2,\ y_3...y_j]$ that could possibly be the order of the system under consideration. Where $u_i$ is the number of past values of the input $i$ and $y_j$ is the number past values of output $j$ that influences the system under consideration.

The dimensionality of an algorithm used for the I-Tier, when identifying a MIMO system with $m$-inputs and $n$-outputs would be $(m + n)$. A swarm of particles (PSO), bee particles (ABC) and antibody population (AIS) are employed to search all possible combination of $[U, Y]$ and identify the order and relative degree of the system. For each order, the network is trained and the model is checked for how well it conforms to the actual system. The mean square error (MSE) is used as the fitness value to check the extent of conformance of the different combinations obtained. The swarm particles (PSO), bee particles (ABC) and antibody population (AIS) are updated such that this value is minimized. This way the different network configurations are evaluated and finally the swarm particles, bee particles and antibody population converge on to the combination that results in the least MSE and which would indicate the order and relative degree of the helicopter system under consideration.

For a MIMO dynamical system with $m$-inputs and $n$-outputs,

*A possible solution: $[u_1,\ u_2,\ u_3...u_i,\ y_1,\ y_2,\ y_3...y_j]$*

$U\ [u_1,\ u_2,\ u_3...u_i]$, $u_i \forall u_i \in Z^+$ *where $i = [1...m]$*

*- number of past history states of each input which has to be included.*

$Y\ [y_1,\ y_2,\ y_3...y_j]$, $y_j \forall y_j \in Z^+$ *where $j = [1...n]$*

*-number of past history states of each output which has to be included.*

Hence the total number of parameters that has to be given as explicit inputs to the network is given by $\sum_{i=1,j=1}^{i=m,j=n}(u_i + y_j)$. $\sum_{i=1}^{i=m} u_i$ - order of the dynamical system and $\sum_{j=1}^{j=n} y_j$ is its relative degree.

This forms the first stage of proposed model wherein the structure of the system is identified, determining all the necessary past inputs and outputs of the system being modelled needed to be fed as explicit inputs to the network.

**B. Tier II**

This module is to train the network identified by Tier I. Here we use a stochastic training algorithm based on PSO/ABC for training the MLP network with TDLs. As indicated earlier these stochastic algorithms have been extensively used for training ANNs and proved to be more efficient than many other gradient based training algorithms (Engelbrecht, P., & Ismail, A., 1999). This can be mainly attributed to stochastic nature of the algorithm which makes it very robust and flexible. In Tier II, the optimum weights are evolved for each network configuration determined by Tier I. The network structure determined by the first level serves as the starting point for the II-Tier.

In this stage, the variable is the weight matrix of the identified network. When applied to Feedforward Neural Network (FNN) training, each particle represents a possible FNN configuration, i.e., its weights. Therefore, each vector has a dimension equal to the number of weights in the FNN.

The unique feature of the current training algorithm, which distinguishes it from the various other similar training algorithms, is the variable/dynamic dimensionality. The dimensionality of the algorithm used for Tier II keeps changing at each iteration, as it depends on the network configuration determined by the outer-tier. Based on the fitness value used, the weights are modified so that the training error is minimized. Hence, finally arriving at a set of weights, that result in the least error for a given network configuration. But from each time to time the number of weights changes as the network configuration changes.

In our study, we use the root mean sum of squared residuals (error) in the training data as the fitness values of the Tier – II PSO/ABC, this serves as a qualitative performance measure of the network learning.

$$rmse = \sqrt{\frac{1}{N_1} \sum_{\in N_1} (\bar{y}(k) - \hat{y}(k))^2} \qquad (9)$$

where $\bar{y}(k)$ is the time value of the output, and $\hat{y}(k)$ is the estimated output of the recurrent neural network. $N_1$ is the number of data points used in the training set. In our case, $\bar{y}(k)$ is the response of helicopter obtained from the flight test. Our objective is to minimize this fitness value and drive it to zero, which then indicates the evolve model to accurately interpolate the data.

The inner loop forms the fitness function evaluation in the outer loop. This is executed every time the fitness function is evaluated in the outer loop. Thus increasing the complexity of the model and making it computationally intensive.

### A. Artificial Bee Colony – Artificial Bee Colony

The model handles helicopter dynamics identification in a two tier architecture using ABC. The first tier ABC is used for evolving the order of the dynamical system under consideration. The second tier ABC is used with the first tier ABC to evolve the optimum weight matrix for the network identified.

In two tier architecture of ABC, the number of bees (n) is initialized. At each time step the randomness amplitude and speed of convergence of each employed and onlooker bees is changed towards its food source (nectar). The random factor prevents the bees getting stuck in local minimae and speed of convergence is used to identify the rate at which employed and onlooker bees exploit a solution. For each of these bees, fitness is evaluated. Our objective is to minimize the fitness value. The Tier II ABC is executed every time the fitness function is evaluated in Tier I ABC.

The selection of the ABC parameters plays an important role in the optimization as the performance of the ABC is quite sensitive to control parameter choices (Karaboga, D., & Basturk, B., 2008). In the current work, there is two tier ABC algorithms being employed, each with different configurations of parameters. The final optimal ABC parameters have been selected after extensive numerical simulations with various combinations. For both cases, optimal refers to the set of ABC parameters which results in fastest convergence along with the most accurate identification of the considered dynamical system. A number of different configurations of parameters have been experimented with trying to achieve a balance between the computational time and the performance. The final optimal set of ABC-ABC parameters for both I and II tier algorithms have been listed in Table 1.

| ABC Parameters | Tier – I ABC | Tier – II ABC |
| --- | --- | --- |
| *Number of Bees* | N = 30 | N = 40 |
| *Randomness amplitude* | $\alpha = 0.45$ | $\alpha = 0.45$ |
| *Speed of convergence* | $\beta = 0.85$ | $\beta = 0.85$ |
| *Learning rate is adaptively generated for each iteration* | $\gamma = [0.5,....,1]$ | $\gamma = [0.5,...,1]$ |
| *Maximum number of iterations* | 50 | 5000 |

**Table 1**. The ABC Parameters

### B. Artificial Immune System – Particle Swarm Optimization

In this model, the order and the relative degree of the system are evolved using AIS in the Tier 1. PSO based learning algorithm is used to train the identified network.

The AIS forms the first stage of the algorithm where in the structure of the system are identified. This network structure serves as starting point of the PSO in the second stage of the algorithm. The antibody (Ab) population is randomly initialized for the AIS. For each of these Ab, fitness is evaluated. The PSO in the Tier II is executed every time the fitness function is evaluated in Tier I

AIS. The dimensionality of PSO keeps changing at each iteration, as it depends on the network configuration determined by the AIS.

After the fitness evaluation of Ab population, various AIS steps of cloning, mutation and re-selection are done to obtain improved set of population. Fitness of this improved population is re-evaluated and new random Ab particles are added to this population for subsequent iteration.

As with the PSO, in the case of AIS also, the parameters play an important role in the optimization as performances of these algorithms are quite sensitive to control parameter choices. The final optimal parameters have been selected after extensive numerical simulations with various combinations. A number of different configurations of parameters have been experimented with trying to achieve a balance between the computational time and the performance. The final set of parameters used for both AIS and PSO are listed in Table 2.

| AIS-PSO Parameters | Tier – I AIS | Tier – II PSO |
|---|---|---|
| Number of clones per antibody | n = 4 | - |
| Hypermutation probability | pm = 0.05 | - |
| Number of Antibodies | P = 4 | - |
| Maximum number of iterations | Max_it = 10 | - |
| Cognitive Learning Rate | - | $C_P = 2.0$ |
| Social Learning Rate | - | $C_g = 2.0$ |
| Inertia factor | - | w = 0.9 |
| Number of Swarm Particles | - | N = 10 |
| Maximum number of iterations | - | 300 |
| End Condition | - | 300 iterations |

**Table 2**. The AIS - PSO Parameters

**C. Particle Swarm Optimization**

This variant, employs a PSO based external algorithm for evolving the order of the system and also a PSO based learning algorithm for the training the evolved network. This model was proposed by Mudigere, D., et.al (2008) and this paper provides a detailed description of the details and characteristics of this model. This is a two-tier PSO architecture, I level PSO for evolving the order of the dynamical system and the II level PSO to evolve the optimum weight matrix for the network identified by the I level PSO.

This model uses the NARX model as the basic substrate with an external PSO shell for evolving the order of the system. The dimensionality of the I-Tier PSO, when identifying a MIMO system with *m*-inputs and *n*-outputs would be *(m + n)*. A swarm of particles are employed to search all possible combination of *[U, Y]* and identify the order and relative degree of the system. In II-Tier PSO, the optimum weights are evolved for each network configuration determined by I-Tier PSO. The network structure determined by the first level PSO serves as the starting point for the II-Tier

PSO. The unique feature of the current training algorithm, which distinguishes it from the various other PSO-based ANN training algorithms, is the variable/dynamic dimensionality. The dimensionality of the II-Tier PSO here keeps changing at each iteration, as it depends on the network configuration determined by the outer-tier PSO.

The selection of these PSO parameters plays an important role in the optimization as the performance of the PSO is quite sensitive to control parameter choices. The final optimal PSO parameters have been selected by after extensive numerical simulation with various combinations. For both cases optimal refers to the set of PSO parameters which results in fastest convergence along with the most accurate identification of the considered dynamical system. A number of different configurations of parameters have been experimented with trying to achieve a balance between the computational time and the performance. The final optimal set of PSO parameters for both for I and II tier algorithms have been listed in Table 3.

| PSO Parameters | Tier – I PSO | Tier – I PSO |
|---|---|---|
| *Cognitive Learning Rate* | $C_P = 2$ | $C_P = 1.85$ |
| *Social Learning Rate* | $C_g = 2$ | $C_g = 1.85$ |
| *Inertia factor* | $w = 0.9$ | $w = 0.9$ |
| *Number of Swarm Particles* | $N = 5$ | $N = 8$ |
| *Maximum number of iterations* | 100 | 1000 |
| *End Condition* (number of iterations without update in the best values) | 25 iterations | 100 iterations |

**Table 3**. The PSO - PSO Parameters

## VI. RESULT AND DISCUSSION

To address the complexity of helicopter dynamics due to strong inter-axis coupling in this paper the longitudinal and lateral states of the helicopter are decoupled and this decoupled model is used for neural network based identification. For simulation purposes in this paper we only consider the Longitudinal Uncoupled Dynamics.

The results of the helicopter dynamics identification is shown in Fig.2 depicting both the actual helicopter response and the predicted response of the model using nature inspired techniques (ABC/AIS/PSO). The actual system response and the predicted response of the system evolved nature inspired techniques agree quite well over the entire range. The performance of the three nature inspired techniques for longitudinal model as shown in Fig. 2. The solid line represents the flight data and dotted line represents network output of nature inspired techniques (dotted blue indicates PSO-NARX, dotted red indicates ABC-NARX and dotted green indicates AIS-NARX). It can be observed that, for the highly nonlinear variation in $q$ and $\theta$, the ABC and PSO appear to give better mapping than the AIS.

A comparison of the NARX model predicted response and the nature inspired techniques predicted response could be seen in Fig.3. It can be seen both ABC and PSO models perform comparatively over the entire range along NARX model. However AIS doesn't give better result for q and θ.

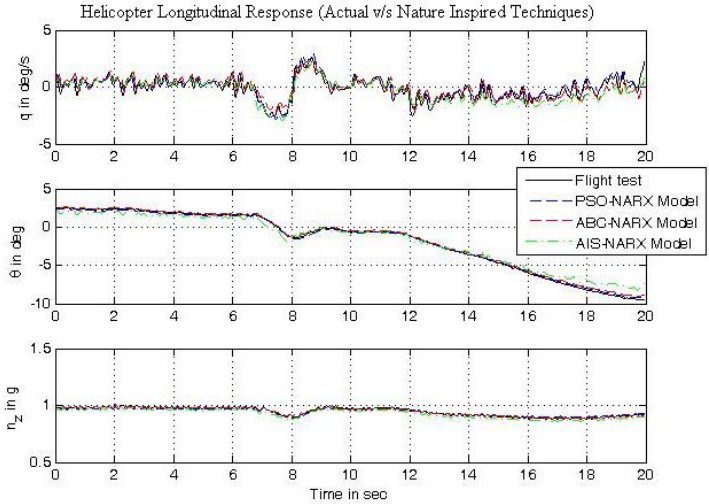
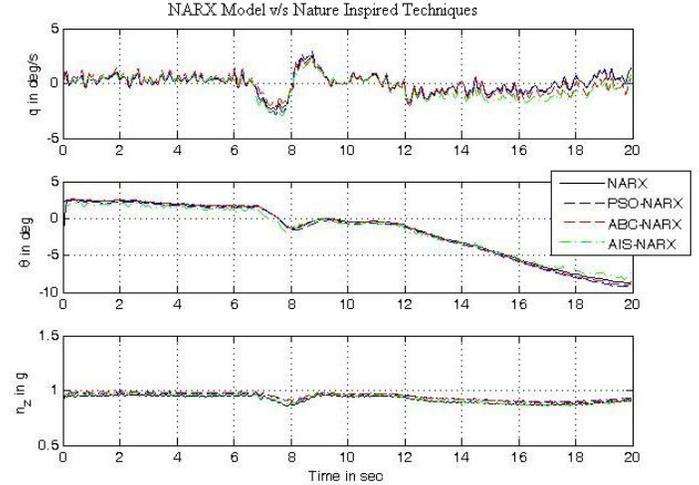

**Fig.2.** Predicted Response and Actual Response for longitudinal cyclic doublet input.

**Fig.3.** Predicted Response of NARX model V/s predicted response of PSO/ABC/AIS-NARX model.

From Table 3, it can be clearly seen that in comparison with other model ABC-NARX evolved model is more efficient and also the training error is significantly reduced. This indicates quite a close correlation between the actual and the predicted response of the helicopter. More or less consistently an order of {1,1} (number of past inputs) and a relative degree of the {1, 2, 0} (number of past outputs) evolved over a number of simulations. But the evolved order and relative degree was not very consistent between the models, a variation of ±1 past value for both the input and output was observed. This could possibly be a short-coming, it has to be further looked into and further means to control this inconsistency has to be introduced. But this variation in not substantial, and more or less the evolved agree well with the actual physics of helicopter dynamics. The evolved configuration can be accounted for as follows; it is quite evident that the third output - normal acceleration ($a_z$), does not have a big influence on the response of the considered longitudinally uncoupled helicopter system. For a longitudinally uncoupled helicopter system, among the outputs the greatest influence on the system response is due to the pitch rate (q). This is clearly identified by the PSO/ABC-NARX and reflected in the evolved configuration. Further, this configuration for longitudinally uncoupled helicopter dynamical system agrees well with the ones reported in literature and seems to be accurate from the aerodynamics/physics of the system.

| Network parameters | # I/p nodes | # O/p nodes | # hidden nodes | Error |
|---|---|---|---|---|
| NARX | 18 | 3 | 20 | 0.281 |
| ABC - NARX | 6 | 3 | 18 | 0.055 |
| AIS - NARX | 6 | 3 | 20 | 0.824 |
| PSO - NARX | 6 | 3 | 28 | 0.068 |

Table 3. Comparison of NARX and PSO/ABC/AIS-NARX model

## VII. CONCLUSIONS

The models proposed in this paper have successfully circumvented the major drawback of the NARX model of having to know system information before hand retaining its other superior characteristics. The use of nature inspired techniques for training the network effectively overcomes the problems associated with back-propagation. Nature inspired techniques with their stochastic means are less likely to get stuck in local minima, making them very robust and flexible. From the results it can be seen the correlation between the predicted response and actual response is satisfactorily accurate in case of ABC-NARX and PSO-NARX. Further in comparison with NARX, PSO-NARX and AIS-NARX, ABC-NARX evolved model is more efficient and also the training error is significantly reduced.

The proposed model is computationally quite resource intensive; this is because of the high degree of complexity involved with the two tier architecture, with having to execute the entire second tier for every objective evaluation of the first tier algorithm. Further work has to be directed in adapting this model for a parallel environment and making it implementation more efficient. Also, in the current work we have restricted the use of this model to identify only the longitudinally uncoupled dynamics of a helicopter. In future work, this model can be used to identify the helicopter dynamics more comprehensively by considering – the laterally uncoupled dynamics, and coupled dynamics also. This would provide more information on the capabilities and limitations of the model. Further these models can be used in varied applications such as flight simulators, modeling, control etc...